\documentclass[twocolumn,amsmath,amssymb,pre,superscriptaddress]{revtex4}
\usepackage{graphicx}
\usepackage{dcolumn}
\usepackage{bm}
\usepackage{subfigure}
\usepackage{booktabs}
\usepackage{url}
\usepackage{color}
\newcounter{one}
\newcommand{\affA}{Department of Computational Intelligence and Systems Science,
Tokyo Institute of Technology, 4259-G5-22, Nagatsuta-cho, Midori-ku, Yokohama, Kanagawa, 226-8502, Japan}

\begin{document}

\title{Persistence of activity on Twitter triggered by a natural disaster: A data analysis}

\author{Tatsuro Kawamoto}
\email{kawamoto.tatsuro@gmail.com}
\affiliation{\affA}
\author{}
\affiliation{\affA}
\date{March 11, 2015}

\begin{abstract}
In this note, we list the results of a simple analysis of a Twitter dataset: the complete dataset of Japanese tweets in the 1-week period after the Great East Japan earthquake, which occurred on March $11$, $2011$.
Our data analysis shows how people reacted to the earthquake on Twitter and how some users went inactive in the long-term.
\end{abstract}

\maketitle

\section{Introduction}
Twitter has attracted significant attention both in academic fields and in industry as a tool that may be useful for sharing and spreading important information during circumstances such as natural disasters.
Many people used Twitter actively in Japan following the March $11$, $2011$ earthquake, which is one example that has led people to tout social media as a useful device.
Although it may be a powerful tool, maintaining such activity is critical.
Suppose that we try to collect or spread information after a natural disaster using Twitter. If the activity of users decays very quickly, we cannot expect substantial response or influence.
The persistence of activity provides another indication. Although the use of hashtags was encouraged after the earthquake so that the users could refer to related information, few users actually used hashtags in their tweets.
It has been argued that this was partly because they were not accustomed to using Twitter.
If such persistence is low, we will face the same problem in a future emergency. 

In this note, we investigate the degree of activity of Twitter users after the earthquake.
We show that many people created accounts and reacted to the information on the earthquake, but a significant fraction of these quickly went inactive afterward.

\section{Dataset and its analysis}
In $2012$, a Japanese project titled \textit{shinsaidata} \cite{shinsaidataWeb} was conducted, and a complete dataset of Twitter activity up to 1 week after the earthquake was distributed.
The project was conducted for research purposes, and the data were accessible for only a limited period.
The dataset contains all tweets posted in Japanese and related information, such as user IDs and time stamps.
Researchers accessing the data agreed to delete the original data when the project ended, and therefore, we no longer possess the data and were only able to retrieve the related statistics.
Based on the user IDs in the dataset, we crawled information using the Twitter API ran based on those IDs.
The analysis presented here was conducted in the fall of $2012$, and we used version $1$ of the API \cite{APIcommand}.

Within the distributed dataset, the total number of unique users who tweeted at least once in Japanese within a week after the earthquake was $N_{\mathrm{total}} = 3,691,599$.
Among them, we confirmed the existence of $N_{\mathrm{data}} = 3,444,326$ ($93.3\%$) accounts using the API, and the remaining users probably deleted their accounts.
Of these users, the number of public users was $N_{\mathrm{public}} = 3,323,342$ ($96.5\%$), whereas we could not search information for private users.
These indicate that most users kept their accounts and that our data analysis is a fairly good estimate of all users in Japan.

\begin{figure}[!t]
\vspace{\baselineskip}
\vspace{\baselineskip}
\hspace{-30pt}
\centering
\includegraphics[width=\columnwidth]{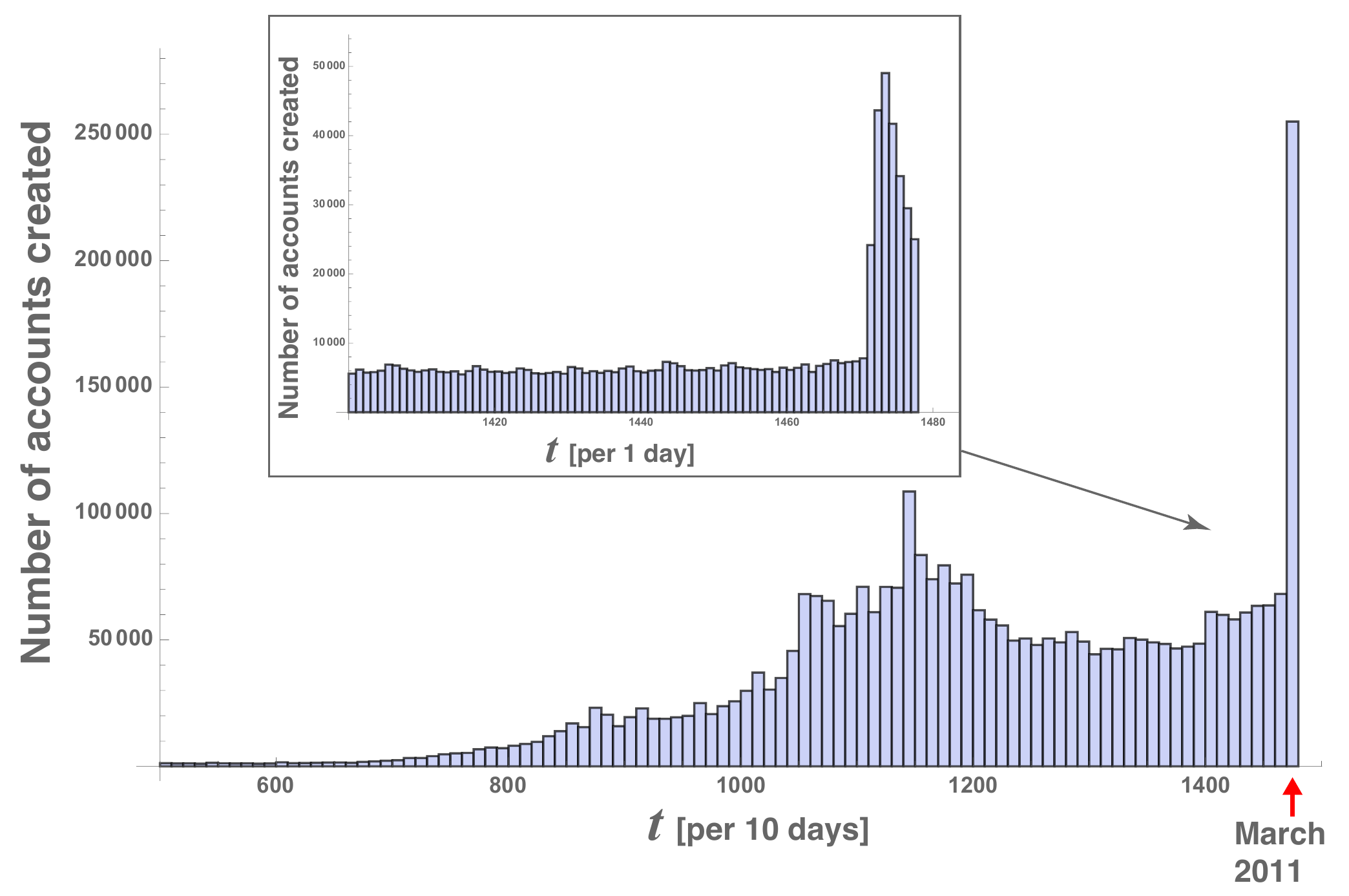}
\caption{Time series of the number of accounts created, where each bin represents a period of $10$ days.
The inset shows the zoomed time series around the day of the earthquake, where each bin represents a period of $1$ day.}
\label{AccountCreation}
\end{figure}

First, we measured the number of people who created accounts after the earthquake.
As shown in Fig.~\ref{AccountCreation}, a significant number of people reacted sensitively to the event and created accounts right after the earthquake.
Specifically, the number of users who joined Twitter within a week after the earthquake was $N_{\mathrm{new}} = 232,818$, which is $6.76\%$ of $N_{\mathrm{data}}$ users, or $222,446$ public users (6.46\% of $N_{\mathrm{data}}$ users).

\begin{figure}[t]
\centering
\includegraphics[width=0.9\columnwidth]{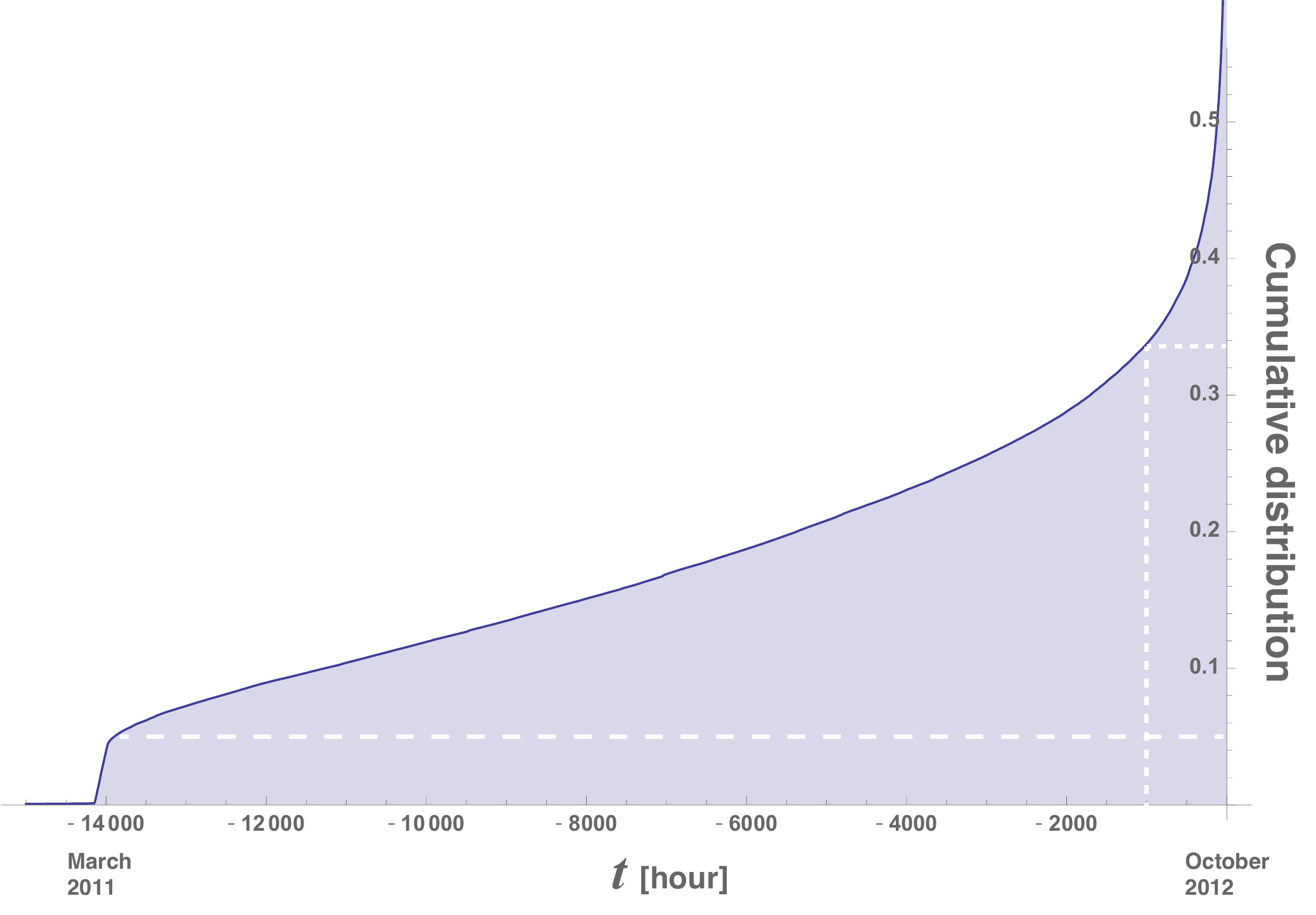}
\caption{Cumulative distribution function of the latest tweet date among $N_{\mathrm{data}}$ users from the day of the earthquake to our measure date.}
\label{CumulativeAll}
\end{figure}

\begin{figure}[h]
\centering
\includegraphics[width=0.9\columnwidth]{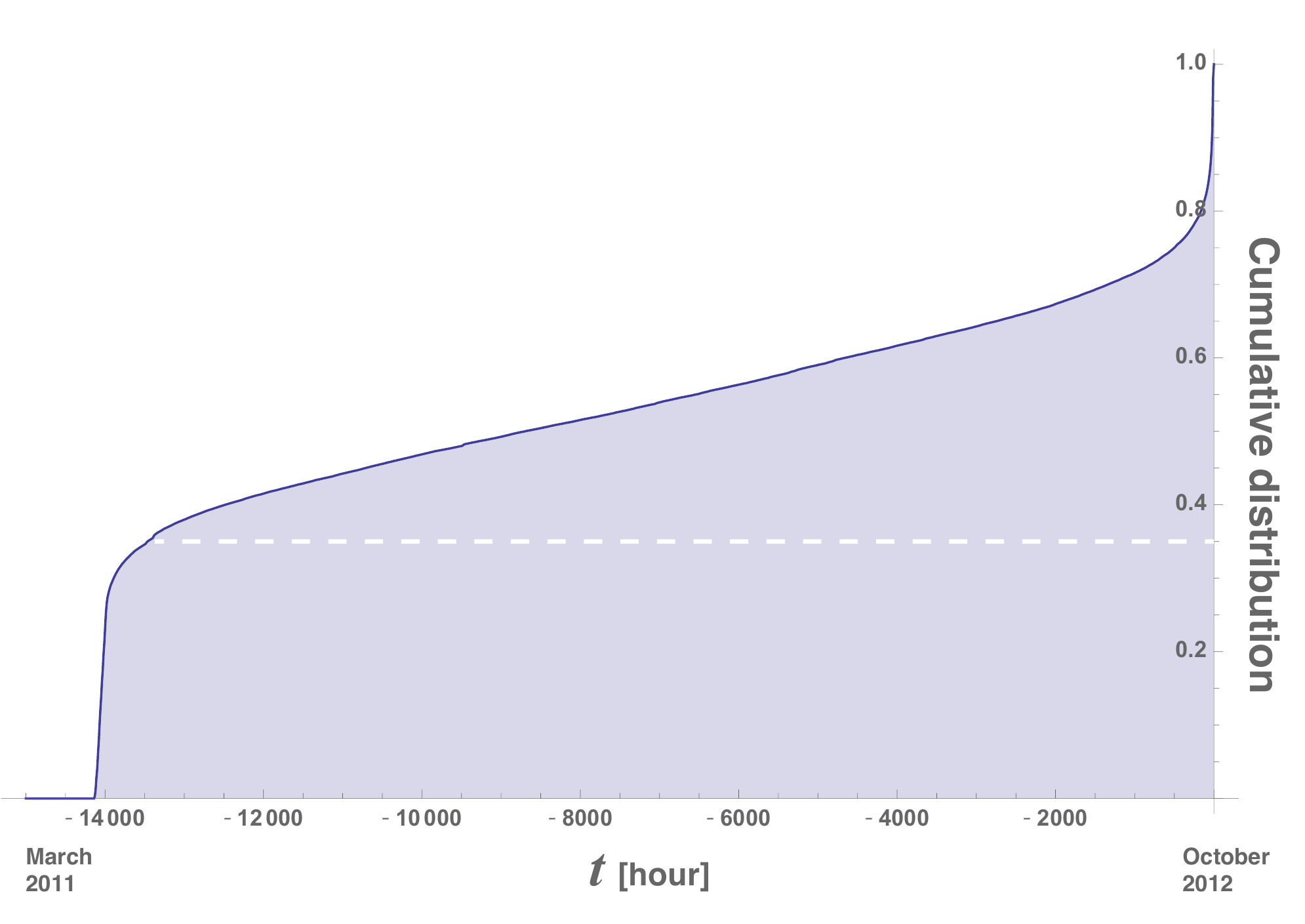}
\caption{Cumulative distribution function of the latest tweet date among $N_{\mathrm{new}}$ users from the day of the earthquake to the measure date.}
\label{CumulativeLow}
\end{figure}

We then plotted the cumulative distribution function of the latest tweet date among $N_{\mathrm{data}}$ users (see Fig.~\ref{CumulativeAll}).
Presented as a reversed J-shaped curve, rapid decay was observed after the earthquake for only $5\%$ of users, who stopped tweeting soon after the earthquake.
On the contrary, if we naively define active users as those who tweeted at least once during a $1,000$-hour period, the fraction of active users was $67\%$ at our measure date.

Finally, we plotted the cumulative distribution function of the latest tweet date among $N_{\mathrm{new}}$ users (see Fig.~\ref{CumulativeLow}).
Roughly $35\%$ stopped tweeting shortly after the earthquake, compared to the small fraction of $N_{\mathrm{data}}$ users who did so.
That is, although a fairly large fraction of all users kept using Twitter, a significant fraction of users who joined shortly after the earthquake practically quit by our measure date ($1.5$ years after the earthquake). Moreover, many of them quit soon after creating their own accounts.

\vspace{\baselineskip}
\vspace{\baselineskip}
\section{Discussion}
Although we defined active users as those who actively tweet, accounts with no tweets do not necessarily mean that they are inactive, as they might be reading Twitter feeds often.
However, there is no way to identify such characteristics on Twitter, and as expected, not all users are active. Our analysis quantitatively shows the degrees of reaction and persistence before and after the March $11$, $2011$ earthquake.

The goal of the workshop \textit{shinsaidata} was to determine whether and how the distributed dataset can be useful for emergencies.
While detailed social network analyses may be possible using the Twitter dataset, we focused on a more fundamental question, because we believe it is relevant to the goal of the project (although not challenging as academic research).
We hope that our simple data analysis contributes as a guideline to estimating the expected reaction and persistence of Twitter users.

\section*{Acknowledgments}
The author thanks Twitter, Inc. for sharing the dataset and the people who organized the workshop \textit{shinsaidata} in $2012$ for providing this opportunity.
The author also thanks Naoaki Okazaki and Yukie Sano for sharing the clean data extracted from the raw dataset and Taro Takaguchi for the useful comments.

\bibliographystyle{apsrev}

\end{document}